\documentclass[10pt,onecolumn]{article}%
\usepackage{amsmath}
\usepackage{amsfonts}
\usepackage{amssymb}
\usepackage[dvips]{graphicx}
\usepackage[margin=3cm]{geometry}%
\newtheorem{theorem}{Theorem}

\newtheorem{proposition}[theorem]{Proposition}

\newenvironment{proof}[1][Proof]{\noindent\textbf{#1.} }{\ \rule{0.5em}{0.5em}}
\begin{document}

\title{Regular quantum graphs}
\author{Simone Severini$^{1}$ and Gregor Tanner$^{2}$\\
{\small $^{1}$ Computer Science Department, University of Bristol, UK }\\
{\small $^{2}$School of Mathematical Sciences, University of Nottingham, UK }}
\maketitle
 
\begin{abstract}
We introduce the concept of regular quantum graphs and 
construct connected quantum graphs with discrete symmetries.
The method is based on a decomposition of the quantum propagator in terms of
permutation matrices which control the way incoming and outgoing channels
at vertex scattering processes are connected.
Symmetry properties of the quantum graph as well as its spectral statistics
depend on the particular choice of permutation matrices, also called connectivity 
matrices, and can now be easily controlled.
The method may find applications in the study of quantum random walks 
networks and may also prove to be useful in analysing universality in spectral
statistics.

\end{abstract}

\section{Introduction}

The study of quantum graphs has become popular in a number of
fields in quantum mechanics ranging from molecular physics and the physics of
disordered systems to quantum chaology and quantum computation (see
\emph{e.g.} \cite{Ku02}). Quantum graphs serve as computationally inexpensive
models with the ability to mimic a variety of features also present in more
realistic quantum systems. For example, the now 20 years old conjecture by
Bohigas, Giannoni and Schmidt (BGS) \cite{BGS84}, stating that the spectral
statistics of quantum systems whose classical limit is chaotic follow those of
random hermitian or unitary matrices in the semiclassical limit is well
reproduced on quantum graphs \cite{KS97, KS99}.

In this paper, we will address two fundamental, but seemingly disconnected
questions related to quantum graphs, namely, we will look at

\begin{itemize}
\item[(1)] \emph{a ways to introduce symmetries on connected quantum graphs}
\end{itemize}

and investigate

\begin{itemize}
\item[(2)] \emph{the degree of complexity or randomness necessary on a
quantum graph to fall within the universal regime of random matrix statistics.} 
\end{itemize}

The first point has hardly been addressed in the context of quantum graphs. 
Symmetries on quantum graphs play an important role in studies on 
quantum random walks considered recently in the context of quantum 
computation (see \emph{e.g.} \cite{Ke03}).  Speed up of mixing-parameters of 
quantum random walks over classical random walks found on certain 
graphs is indeed related to interference effects due to symmetries in the 
quantum propagation. We will suggest a method for imposing a large class of symmetries 
on certain types of graphs which has potential applications in the design 
of effective quantum random walks. 

It is furthermore expected that symmetries on graphs will have a profound 
influence on the statistical properties of spectra of quantum graphs.
The existence of discrete symmetries and associated \textquotedblleft good 
quantum numbers\textquotedblright\ on connected quantum graphs is 
expected to lead to deviations from random matrix results. 

The second point addresses the range of validity of the BGS - conjecture. It is
widely believed that the spectra of unitary propagators on quantum graphs follow 
random matrix statistics if the correlation exponents of an underlying stochastic 
dynamics are bound away from zero in the limit of large graphs sizes and the length of
the arcs of the graph are incommensurate \cite{Ta01}. We will argue here that
the last condition can be considerably relaxed and that, in context of regular 
graphs, the existence or absence of random matrix statistics is 
related to the commutativity properties of certain sets of 
\emph{connectivity matrices} to be defined in detail later. Similar results
for the spectra of Laplacians of regular graphs have been reported in 
\cite{JMRR99}. A related discussion of spectra of adjacency matrices of  
Cayley graphs of certain groups can be found in \cite{Te03}. \\

We start by briefly reviewing the notion of a quantum graph. A quantum graph 
is given by an underlying graph $G$ and a set of local scattering matrices at
the vertices as well as a set of arc lengths. A (finite) \emph{directed graph} 
or \emph{digraph} consists of a finite set of \emph{vertices} and a set of 
ordered pairs of vertices called \emph{arcs}. We denote by $V^{G}$ and 
$E^{G}$ the set of vertices and 
the set of arcs of the digraph $G$, respectively. Given an ordering of the 
vertices, the \emph{adjacency matrix} of a digraph $G$ on $n$ vertices, 
denoted by $A^{G}$, is the $(n\times n)$ $\left(  0,1\right)  $-matrix where 
the $ij$-th element is defined by
\[
A_{ij}^{G}:=\left\{
\begin{tabular}
[c]{ll}%
$1$ & if $(i,j)\in E^{G},$\\
$0$ & otherwise.
\end{tabular}
\right.
\]
An \emph{undirected graph} (for short, \emph{graph}) is a digraph whose
adjacency matrix is symmetric. The \emph{line digraph} of a digraph $G$,
denoted by $LG$, is defined as follows, (see \emph{e.g.} \cite{BG01}):
$V^{LG}=E^{G}$ and, given $(h,i),(j,k)\in E^{G}$, the ordered pair 
$((h,i),(j,k))\in
 E^{LG}$ if and only if $i=j$. 

A \emph{quantum graph} associated with a digraph
$G$ on $n$ vertices may then be defined in terms of a set of unitary vertex
scattering matrices $\sigma^{(j)}$ on vertices $j=1,\ldots n$ and a set of 
arc-lengths $L_{(i,j)}$ defined for every arc $(i,j)\in E^{G}$. Waves 
propagate freely along the directed arcs, transitions between incoming and
outgoing waves at a given vertex $j$ are described by the scattering matrix
$\sigma^{(j)}$. The two sets specify a unitary propagator of dimension 
$n_{E}=|E^{G}|$ defining transitions between arcs 
$(i,j),(i^{\prime},j^{\prime})\in E^{G}$ 
which has the form \cite{KS97}
\[
\begin{tabular}
[c]{ccc}
$S^{G}=D\, V$ & with & $D_{(i,j)(i^{\prime},j^{\prime})}=
\delta _{i,i^{\prime}}\, \delta _{j,j^{\prime}}\, e^{\mathrm{i}kL_{(i,j)}},$%
\end{tabular}
\]
where $k$ is the wave number and
\[
\begin{tabular}
[c]{ccc}
$V_{(i,j)(i^{\prime},j^{\prime})}=A_{(i,j)(i^{\prime},j^{\prime})}^{LG}%
\sigma_{ij^{\prime}}^{(j)}$ & with & $A_{(i,j)(i^{\prime},j^{\prime})}%
^{LG}=\delta_{j,i^{\prime}}$ $.$%
\end{tabular}
\]
The local scattering matrices $\sigma^{(i)}$ depend on the boundary conditions
and local potentials at the vertex $i$ which we do not want to specify here 
any further. For our purpose, we may regard the $\sigma^{(i)}$'s as arbitrary
unitaries. Let $d_{i}^{-}$ and $d_{i}^{+}$ be the number of incoming and
outgoing arcs of a vertex $i$, respectively. A sufficient and necessary 
condition for a digraph $G$ to be quantisable in the way above is then, that
for every vertex $i\in V^{G}$, $d_{i}^{+}=d_{i}^{-} =
d_{i}=\dim\sigma^{(i)}$ \cite{PTZ03, S03}. This means in particular that if 
$G$ is an undirected graph then it is quantisable. 
 
The "classical"\ dynamics corresponding to a quantum graph defined by a unitary
propagator $S^{G}$ is given by a stochastic process with transition matrix $T$ 
\[
T_{ij}=|S_{ij}^{G}|^{2}=|V_{ij}|^{2} \, .
\]
Note that both the quantum mechanics as well as the associated stochastic
dynamics relate to transitions between arcs and is thus defined on the line 
digraph of $G$.\newline
 
The paper is organised as follows. In Section \ref{sec:reg-graph}, we will
introduce the notion of regular quantum graphs and discuss a factorisation 
of the propagator in terms of connectivity matrices for a special 
class of such graphs. In Section \ref{sec:sym-reg}, we relate the 
existence or absence of symmetries on a connected regular quantum graph to 
properties of the connectivity matrices. We discuss some specific examples for
completely connected graphs including statistical properties of the spectra
in Section \ref{sec:examples}. In Section \ref{sec:rmt} we show numerically 
that by inscribing a {\em single} ($2\times 2)$ unitary matrix into 
a large regular quantum graph one still obtains random matrix 
statistics despite huge degeneracy in the set of arc lengths and scattering
matrices for a generic choice of connectivity matrices. 
 
\section{Regular quantum graphs\label{sec:reg-graph}}
 
We will implement symmetries on quantum graphs for which 
the wave dynamics at the vertices of the digraph is "locally 
indistinguishable" when going from one vertex to the next. We will 
restrict ourselves to wave dynamics on $d$-regular digraphs. Recall 
that a digraph $G$ is said to be $d$
\emph{-regular} if, for every vertex $i\in G$, $d_{i}^{+}=d_{i}^{-}=d$ and thus
$|E^{G}|=nd$. Extending the concept of local
indistinguishableness to quantum graphs, we will consider quantum graphs on
$d$-regular digraphs with local $(d\times d)$ scattering matrices 
$\sigma^{(i)}$ and set of outgoing arc lengths $L_{(i,j)}$ being identical 
at every vertex $i$ up to permutations of the incoming or outgoing channels. 
That is, there are $(d\times d)$ unitary matrices $\sigma$ and $D(k)$ 
with $D_{ij}(k)=\delta_{i,j}\, \exp({\mathrm{i}k L_{i}})$ and local permutation 
matrices $q^{(i)},p^{(i)}$, such that
\[
\begin{tabular}
[c]{ccc}%
$\sigma^{(i)}=p^{(i)}\sigma q^{(i)}$ & and & $D^{(i)}(k)=p^{(i)}D(k)\left(
p^{(i)}\right)  ^{-1}$ $.$
\end{tabular}
\ \ \ \
\]
Combining the local matrices $\sigma$ and $D(k)$ to a single matrix 
$C(k) = D(k)\sigma$, we obtain
\begin{equation}
C^{(i)}(k)=D^{(i)}(k)\sigma^{(i)}=p^{(i)}Cq^{(i)} \, .
\label{coin}
\end{equation}
We call a quantum graph with these properties a \emph{regular quantum graph}.
The matrix $C$ is called the \emph{coin} in the context of quantum random 
walks on graphs \cite{Ke03}.\newline

We denote by $J_{n}$ the $(n\times n)$ matrix with all elements being equal to 1 and 
$I_{n}$ is the identity matrix. The following observation will be useful in 
what follows, (see also \cite{S03a}):
 
\begin{proposition}
\label{proadjLG} Let $A^{G}$ be the adjacency matrix of a $d$-regular digraph
$G$. The adjacency matrix of $LG$ has up to reordering the arcs the form
\begin{equation}
A^{LG}=\left(  \bigoplus_{i=1}^{d}\rho_{i}\right)  \cdot\left(  J_{d}\otimes
I_{n}\right)  \text{ }\, , \label{adj-LG}
\end{equation}
where the matrices $\rho_{i}$ of dimension $n$ have entries 0 or 1 and 
represent discrete functions on the vertex set, that is 
$j\in V^G \rightarrow\rho_{i}(j) \in V^G, \forall j \in V^G$ and in addition
\begin{equation}
\sum_{i=1}^{d}\rho_{i}=A^{G} \, .
\label{cond-rho}
\end{equation}
 
\end{proposition}

\begin{figure}
  \begin{center} \includegraphics[height=3.5cm]{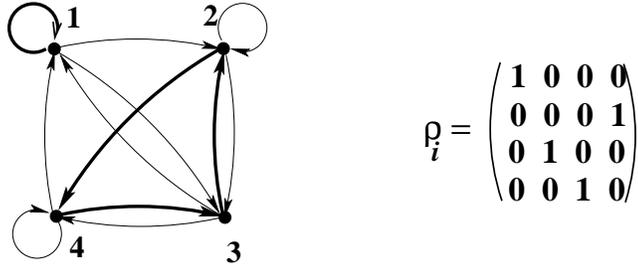} \end{center}
  \caption{A 3-regular graph of size 4 together with a possible 
   connectivity matrix relating an outgoing arc to every vertex.} 
  \label{Fig:graph} 
\end{figure}

\textbf{Remark:} A given matrix $\rho_i$ assigns to every vertex $j$ a specific 
arc $(j,\rho(j)$, see Fig.~\ref{Fig:graph}. Note that, for $d>1$, the choice of 
matrices $\rho_{i}$ is not unique and that the $\rho_{i}$'s do not need to be 
invertible. Different
sets of $\rho_{i}$'s fulfilling the conditions in the proposition give rise to
adjacency matrices of the line digraph which are equivalent up to permutations
in the ordering of the arcs in (\ref{arc-ord}).\\ 
 
\begin{proof}
Condition \ref{cond-rho} ensures that $(j,\rho_{i}(j))\in E^{G}$ for every $i$
and $j$; writing out Eqn.\ (\ref{adj-LG}), we obtain
\[
\bigoplus_{i=1}^{d}\rho_{i}=\left[
\begin{array}
[c]{cccc}
\rho_{1} & 0 & \cdots & 0\\
0 & \rho_{2} & \cdots & 0\\
\vdots & \vdots & \ddots & \vdots\\
0 & 0 & \cdots & \rho_{d}%
\end{array}
\right], \quad  \mbox{and thus} \quad 
\left(  \bigoplus_{i=1}^{d}\rho_{i}\right) \cdot
\left(J_{d}\otimes I_{n}\right) = \left[
\begin{array}
[c]{cccc}%
\rho_{1} & \rho_{1} & \cdots & \rho_{1}\\
\rho_{2} & \rho_{2} & \cdots & \rho_{2}\\
\vdots & \vdots & \ddots & \vdots\\
\rho_{d} & \rho_{d} & \cdots & \rho_{d}%
\end{array}
\right] \, .
\]
The choice of matrices $\rho_i$ fixes now a certain ordering of the arcs;
Ordering the arcs according to
\begin{equation}\label{arc-ord}
(1,\rho_{1}(1)),(2,\rho_{1}(2)),\ldots,(n,\rho_{1}(n)),
(1,\rho_{2}(1)),\ldots,(n,\rho_{2}(n)),\ldots,
(1,\rho_{d}(1)),\ldots,(n,\rho_{d}(n))\, ,
\end{equation}
one deduces that non-zero matrix elements of $A^{LG}$ as defined in 
(\ref{adj-LG}) refer to transitions
\[
(i\rho_{j}(i))\rightarrow(\rho_{j}(i)\rho_{k}(\rho_{j}(i)))\text{ },
\]
for every $j,k=1,\ldots d$ and every $i=1,\ldots n$, which are exactly the
allowed transition in the line digraph of $G$.\newline
\end{proof}

We pointed out in the introduction that the wave propagation on a quantum
graph actually lives on the line digraph of the underlying digraph. 
Generalising (\ref{adj-LG}) to describe unitary propagation on digraphs, 
we write
\begin{equation}
S^{G}=\left(  \bigoplus_{i=1}^{d}\rho_{i}\right)  \cdot\left(  C\otimes
I_{n}\right)  =\left[
\begin{array}
[c]{cccc}
C_{11}\,\rho_{1} & C_{12}\,\rho_{1} & \cdots & C_{1d}\,\rho_{1}\\
C_{21}\,\rho_{2} & C_{22}\,\rho_{2} & \cdots & C_{2d}\,\rho_{2}\\
\vdots & \vdots & \ddots & \vdots\\
C_{d1}\,\rho_{d} & C_{d2}\,\rho_{d} & \cdots & C_{dd}\,\rho_{d}%
\end{array}
\right]  , \label{qu-LG}%
\end{equation}
with $C$ being the unitary $(d\times d)$ coin and the matrices $\rho_{i}$ 
fulfil the condition (\ref{cond-rho}). We have to add the additional 
constraints here that the $\rho_{i}$'s are invertible, that is, that they are 
permutation matrices. The condition is necessary to ensure that $S^{G}$ is unitary. 
We will refer to the permutations $\rho_{i}$ as the 
\emph{connectivity matrices} in what follows, see \ref{Fig:graph}.
$S^{G}$ satisfies all the properties of a regular quantum graph as defined
above. The matrix $C$ is in particular the coin from which the local scattering
matrices $C^{(j)}$ at vertices $j$ can be deduced. One obtains
\[
C_{kl}=C_{\rho_{k}^{-1}(j)\rho_{l}(j)}^{(j)}\text{ }.
\]
The connectivity matrices $\rho_{k}$ and $\rho_{l}$ specify thus the pair of 
arcs related through the transition $C_{kl}$ at a given vertex $j$. 

\textbf{Remark:} In contrast to (\ref{adj-LG}) where different decomposition 
of $A^{G}$ of the form (\ref{cond-rho}) lead to equivalent adjacency matrices
(up to reordering the arcs), this is no longer the case for (\ref{qu-LG}).
Different sets of connectivity matrices lead here to different regular quantum 
graphs which may have very different spectral properties as will be discussed 
in the next section. 

\textbf{Remark:} Note that not all regular quantum graphs can be written in the 
form (\ref{qu-LG}). Any pair of permutation matrices $P$ and $Q$ leaving the 
adjacency matrix of a line graph $A^{LG}$ of a $d$-regular graph $G$ invariant, 
that is $Q A^{LG} P = A^{LG}$, transform an associate propagator of a regular 
graph, $S^G$, into a propagator of a $d$-regular, albeit different, quantum 
graph $\tilde{S}^{G}=P\,S^{G}\, Q$.  If $S^{G}$ is of the form (\ref{qu-LG}),
one easily finds permutations $P$ and $Q$ such that $\tilde{S}^{G}$ is not 
of this form.\newline

So far we have considered general regular digraphs. In the special case 
where the underlying graph is undirected it is natural to consider 
associated time-reversal symmetric regular quantum graphs; that is,
regular quantum graphs for which for every (wave)-paths there 
exists an equivalent time-reversed paths undergoing the same transitions.
A time-reversal symmetric unitary propagator of the form (\ref{qu-LG}) for 
an undirected regular graph can be constructed by choosing 
symmetric coin and connectivity matrices, that is,
\[
\begin{tabular}
[c]{ccccc}
$C=C^{\intercal}$ & and & $\rho_{i}=\rho_{i}^{\intercal},$ &  & for every
$i=1,\ldots, d$ $.$
\end{tabular} 
\ 
\] 
Note, that the symmetry conditions for the connectivity matrices severely limit 
the choice of possible graphs and decompositions.

\section{Symmetries on regular quantum graphs and spectral
decompositions\label{sec:sym-reg}}
 
We note first that if a regular quantum graph can be written in the form
(\ref{qu-LG}) and there exists an invertible $(n\times n)$ matrix $\pi$ such
that
\begin{equation}%
\begin{tabular}
[c]{ccc}%
$\lbrack\pi,\rho_{i}]=0$ &  & for every $i=1,\ldots d.$%
\end{tabular}
\ \label{comm}%
\end{equation}
then
\begin{equation}%
\begin{tabular}
[c]{ccc}%
$\lbrack P,S^{G}]=0$ & with & $P=\left(  I_{d}\otimes\pi\right)  $%
\end{tabular}
\ \label{comm1}%
\end{equation}
independently of the choice of the coin $C$. The result follows immediately
from
\[
\lbrack\left(  C\otimes I_{n}\right)  ,P]=0\text{ }.
\]
It is obvious that the condition (\ref{comm}) implies
$\lbrack\pi,A^{G}]=0\text{ }$.
 
The above property enables us to study certain symmetries of quantum graphs in terms
of the symmetries of the connectivity matrices only. Given a $d$-regular 
digraph $G$ we can in general find many sets of connectivity matrices which sum
up to $A^{G}$ and which may have very different symmetry properties. Or if 
one is interested in quantum graphs with specific symmetries one may start 
from a set of connectivity matrices $\rho_{i}$ in order to construct quantum 
graphs with desired properties. We consider various scenarios here and give 
some specific examples in the next section.
 
\subsection{The abelian case: $[\rho_{i}, \rho_{j}]=0$}

In the special case when all connectivity matrices commute, every 
$\rho_i$ acts as a symmetry $pi$. The spectrum of $S^{G}$ can then 
be decomposed into $n$ sub-spectra of dimension $d$.

\begin{proposition}
\label{prop1} Let $S^{G}$ be of the form
\[
S^{G}=\left(\bigoplus_{i=1}^{d}\rho_{i}\right)  \cdot\left(  C\otimes
I_{n}\right)\text{ },
\]
where the $(n\times n)$ connectivity matrices $\rho_{i}$ fulfil $[\rho_{i}
,\rho_{j}]=0$, $A_{G}=\sum_{i=1}^{d}\rho_{i}$
is the adjacency matrix of a d-regular digraph and $C$ is a $(d\times d)$ 
unitary matrix. Let $u$ be the $(n\times n)$ unitary matrix 
simultaneously diagonalising the $\rho_{i}$'s, that is,
\[
u^{\dagger}\rho_{i}u = 
\bigoplus_{m=1}^n \, e^{\mathrm{i}\varphi_{m}^{i}} \quad i=1,\ldots, d
\]
and $\varphi_{m}^{i}$ is the $m$-th eigenphase of the connectivity matrix
$\rho_{i}$ where the order is determined by the transformation $u$. The
spectrum of $S^{G}$, $sp\left(  S^{G}\right)  $, is then 
\[
sp\left(  S^{G}\right)  =sp\left(  S_{1}^{G}\right)  \uplus sp\left(
S_{2}^{G}\right)  \uplus\cdots\uplus sp\left(  S_{n}^{G}\right)  \text{ },
\]
where
\begin{equation}
S_{m}^{G}= \left(\bigoplus_{i=1}^d \, e^{\mathrm{i}\varphi_{m}^{i}}\right) 
\cdot C \, . 
 \label{sub-ab}
\end{equation}
\end{proposition}

\begin{proof}
Define $U=\left(  I_{d}\otimes u\right)$ and note that
\[
\lbrack U,\left(  C\otimes I_{n}\right)  ]=0.
\]
Thus
\[
U^{\dagger}S^{G}U=\left(\bigoplus_{i=1}^{d} 
\bigoplus_{m=1}^n \, e^{\mathrm{i}\varphi_{m}^{i}}  \right)
\cdot\left(  C\otimes I_{n}\right)  \,.
\]
There exist permutation matrices $P$ such that
$P^{T}\left(  C\otimes I_{n}\right)  P=\left(  I_{n}\otimes C\right)$
and thus
\[
P^{T}\,U^{\dagger}S^{G}\,U\,P=\left(  \bigoplus_{i=1}^{n}S_{m}^{G}\right)
\]
is indeed block-diagonal of the form stated in the proposition.\newline
\end{proof}

\textbf{Remark:} Note that the decomposition is independent of the coin
$C$.\newline
 
It can be shown \cite{ST04} that a set of commuting connectivity matrices of 
a connected graph $G$ always form a subset of the regular (permutation)
representation of an abelian group and the underlying symmetry of the
corresponding quantum graph is given by that group. (The commutativity of the
$\rho_{i}$'s does in fact imply that $G$ is a Cayley digraph of an abelian
group; it must therefore have the form of a discretised torus). The sub-spectra 
obtained from $S_{m}^{G}$ may then be characterised in terms of the eigenbasis of the
generators of the abelian group represented by the connectivity matrices. Let
$a_{1},\ldots a_{r}$ be the generators of such an abelian group,
\[
\begin{tabular}
[c]{cccc}
with $a_{i}^{n_{i}}=id$ & and $\prod_{i=1}^{r}n_{i}=n,$ & where & $n_{i}
\geq2.$
\end{tabular}
\]
The eigenbasis of the connectivity matrices may then be written in Dirac 
notation as $|m_{1}\ldots m_{r}\rangle$ with $m_{i}=1,\ldots n_{i}$ and the 
sub-spectra obtained from (\ref{sub-ab}) are characterised by a set of $r$ 
``quantum numbers'' $S_{m_{1},\ldots,m_{r}}^{G}$. Such a regular quantum graph 
is thus a
discretized version of a quantum systems whose underlying classical dynamics
has $r$-integrals of motion in involution. Some additional degrees of freedom
are represented by the coin $C$ which may or may not be related to classical 
chaotic dynamics depending on the properties of $C$ and the group. 
 
\subsection{Partial symmetries: $[\pi,\rho_{i}] = 0$, 
but $[\rho_{i},\rho_{j}] \ne0$} \label{sec:sec-part} 
Next we consider the case that a symmetry $\pi$ exists
with $[\pi,\rho_{i}]=0$ for all $i\in V^{G}$, but $[\rho_{i},\rho_{j}]\neq0$
for some $i,j = 1,\ldots, d$. That implies that $\pi$ has degenerate eigenvalues;
$\pi$ could for example represent a $C_{2}$ symmetry of the graph, that
is, $\pi^{2}=I_{n}$ with eigenvalues $\pm 1$ only. 

Let us assume that $\pi$ has $r<n$ distinct eigenvalues 
$\lambda_i$, $i=1,\ldots, r$, each with multiplicity $n_{i}$ with
\[
\sum_{i=1}^{r}n_{i}=n.
\]
Let $u$ be a unitary matrix diagonalising $\pi$ in the form
\[
u^{\dagger}\pi u=\bigoplus_{i=1}^{r}\Lambda_i I_{n_{i}} \, ;
\]
$u$ then brings $\rho_{i}$ into block-diagonal form, that is,
\[
u^{\dagger}\rho_{i}u=\bigoplus_{j=1}^{r}\tilde{\rho}_{i}^{(j)}
\]
with $\dim\tilde{\rho_{i}}^{(j)}=n_{j}$. The spectrum of $S^G$ is now 
decomposed in the following way:

\begin{proposition}
\label{prop2}Let $S^{G}$ be of the form (\ref{qu-LG}) and the matrices
$\rho_{i}$, $\pi$ have the properties as described above; $sp\left(
S^{G}\right)  $ is then of the form
\[
sp\left(  S^{G}\right)  =sp\left(  S_{1}^{G}\right)  \uplus sp\left(
S_{2}^{G}\right)  \uplus\cdots\uplus sp\left(  S_{r}^{G}\right)  \text{ },
\]
with
\begin{equation}%
\begin{tabular}
[c]{ccc}%
$S_{m}^{G}=\left(  \bigoplus_{i=1}^{d}\tilde{\rho}_{i}^{(m)}\right)
\cdot\left(  C\otimes I_{n_{m}}\right)  $ & where & $m=1,\ldots, r$ $.$%
\end{tabular}
\ \label{sub-pi}%
\end{equation}

\end{proposition}
 
The proof goes along the line of the proof of Proposition \ref{prop1}. 

The decomposition is again independent of the coin $C$, but the sub-spectra
are now of dimension
\[
\dim S_{m}^{G}=d\, n_{m}\text{ }.
\]
There is a trivial symmetry independent of the particular choice of the
$\rho_{i}$'s related to the fact that every permutation matrix has an
eigenvalue 1 with corresponding eigenvector $(1,1,\ldots,1)^{T}$. The symmetry
$\pi$ in question has the form
\[
\pi_{ij}=\frac{2}{n}-\delta_{i,j}
\]
having two distinct eigenvalues $\pm1$ and the eigenvalue $-1$ has geometric
multiplicity $n-1$. As a consequence any $S^{G}$ can be block-diagonalised
containing $C$ as an $(d\times d)$ block, and thus
\[
sp(C)\subset sp(S^{G})\text{ }.
\]

\section{Some examples for $A^{G}=J_{n}$\label{sec:examples}}

When constructing particular examples, it is useful to start with the
completely symmetric graph, namely that of a \emph{fully connected graph}.
This graph, also called the \emph{complete graph}, has adjacency matrix $J_{n}$.
As $[P,J_{n}]=0$ for every permutation matrix $P$ of size $n$, we may indeed
construct regular quantum graphs of degree $d=n$ with whatever finite symmetry
we want. In addition, we can make use of the fact that if $\Gamma$ is a finite
group of order $n$ and the $(n\times n) $ permutation matrices $\rho_{i}$
form a regular representation of $\Gamma$ then
\begin{equation}
\sum_{i=1}^{n}\rho_{i}=J_{n}\,. \label{cond-Jd}
\end{equation}
We can thus implement the group properties of any finite group on a regular
quantum graph by choosing the regular representations of that group as the 
connectivity matrices. 
In what follows we will consider various decompositions of $J_{n}$ and see
how they effect spectral properties like level statistics.

\subsection{The cyclic group $\mathbb{Z}_{n}$}

The simplest abelian group is the cyclic group $\mathbb{Z}_{n}$. The regular
representations $\rho_{i}$ are of the form
\[
\begin{tabular}
[c]{ccccc}
$(\rho_{j})_{kl}=\delta_{k,(l+j)\operatorname{mod}n}$ & with eigenvalues &
$\chi^j_{m}=e^{2\pi\mathrm{i} \frac{jm}{n}},$ & where & $j,m=1,\ldots, n\,.$%
\end{tabular}
\
\]
Here, $\rho_{j}=\left(  \rho_{1}\right)^{j}$ and $\rho_{n}=I_{n}$. In order
to construct regular quantum graphs with circular symmetry independent of the 
coin $C$ we use the regular representation of $\mathbb{Z}_{n}$ as connectivity 
matrices. The spectrum of the quantum graph can then be decomposed into the sub-spectra
given by
\[
S_{m}^{G}= 
\left( \bigoplus_{j=1}^n e^{2\pi\mathrm{i} \frac{j m}{n}} \right) \cdot C \, .
\]
The eigenvalues are characterised in terms of two quantum numbers, an `angular
momentum' quantum number $m$ and a second quantum number $r$, say, counting
the eigenvalues in each $m$ manifold. If the spectra for different $m$ are
uncorrelated, one expects Poisson statistics of the total spectrum in the
limit $n\rightarrow\infty$.
 
\begin{figure}
  \begin{center} \includegraphics[height=19cm]{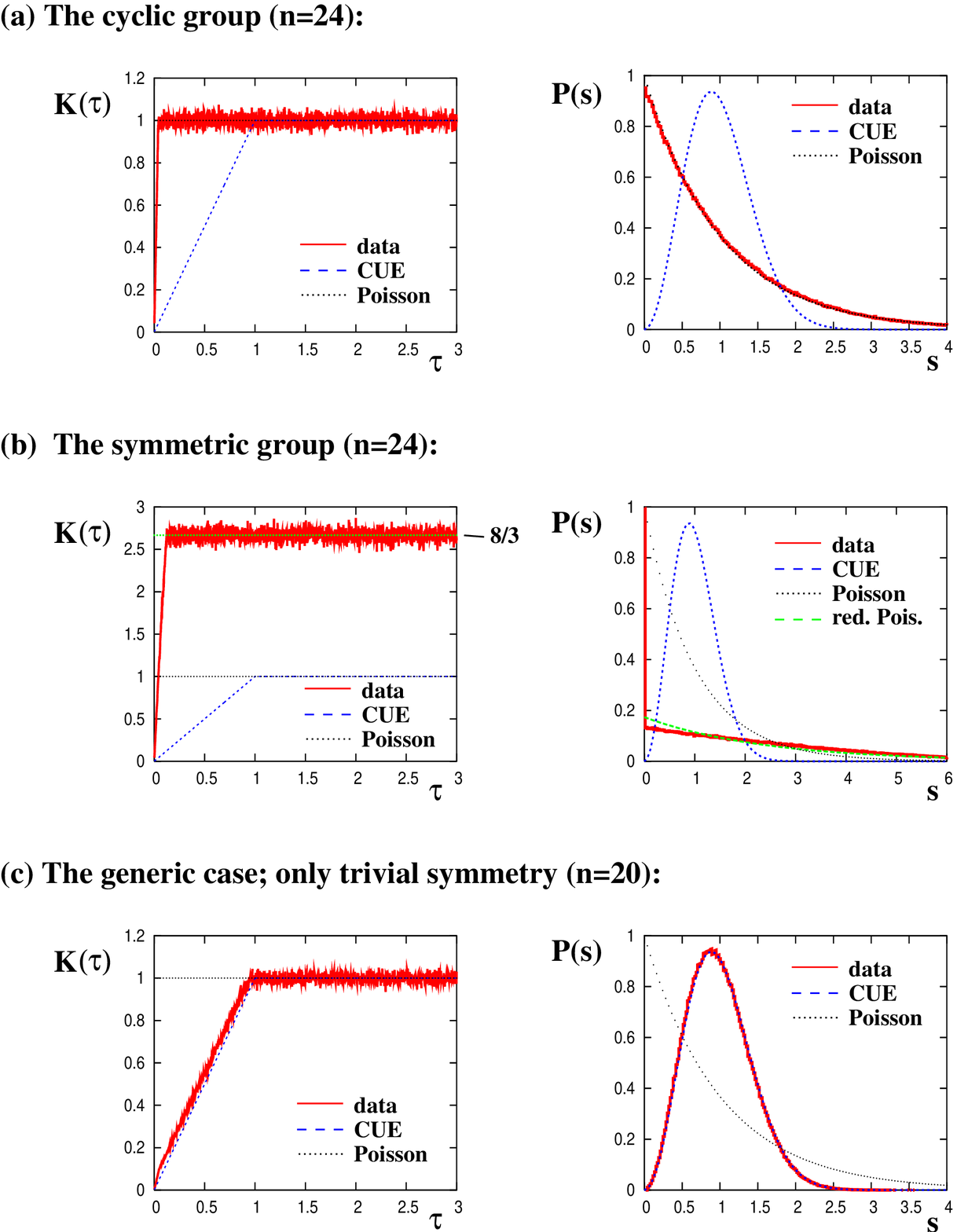} \end{center}
  \caption{Formfactor $K(\tau)$ and nearest neighbour spacing
  distribution $P(s)$ for (a) $\rho_i$'s are the regular
  representation of the cyclic group $\mathbb{Z}_{24}$; (b)
  $\rho_i$'s represent the symmetric group $S_4$; (c) a generic set
  $\rho_i$'s without non-trivial symmetries.
  The dashed curve in (b) labelled ''red. Poisson''
  corresponds to a distribution of degenerate levels being
  Poisson distributed otherwise.} 
  \label{fig:stat} 
\end{figure}

Figure \ref{fig:stat}a) shows spectral properties of $S^{G}$ with $n=24$, 
that is, $\dim S^{G}=576$. We plot here the nearest neighbour spacing (NNS) 
distribution $P(s)$ and the form factor $K(\tau)$, the Fourier transform of 
the two-point correlation function. The coin is of the form (\ref{coin}) where 
the local scattering matrix $\sigma$ is taken randomly from a $CUE$ -  ensemble, 
but then fixed, and the arc lengths entering the diagonal matrix $D$ are chosen
independently and identically distributed in $[0,1]$, but then fixed. The
average is taken over the wavelength $k$ \cite{Ta01}. \footnote{Alternatively, one
can consider the spectrum being the "resonances" of the quantum graph present
at wave-numbers fulfilling
\[
\det\left(1-S^{G}(k)\right)  =0\,.
\]
Both approaches are equivalent under very general conditions \cite{BK99}.} The 
numerical results shown in Fig.~\ref{fig:stat}a)  indeed suggest Poisson-statistics 
apart form deviations in the form factor on scales $\tau\leq1/n$ due to the 
`chaotic nature' of the coin.

\subsection{The non-abelian case: the symmetric group $S_{4}$}
 
Next, we consider a specific example of a non-abelian group, namely the
symmetric group $S_{4}$ with $n=24$ elements; we will discuss spectral
properties of general groups elsewhere \cite{ST04}. The regular representation
of $S_{4}$ can be decomposed in terms of its irreducible representations (for
short \emph{irreps}); each $\rho_{i}$ contains each $d$-dimensional
irrep exactly $d$ times. The group $S_{4}$ has 2 one-dimensional, 1
two-dimensional and 2 three-dimensional irreps, such that
\[
2\cdot1^{1}+1\cdot2^{2}+2\cdot3^{3}=24.
\]
Denote the irreps of the group element $i\in S_{4}$ as
\[
\tilde{\rho}_{i}^{(1,1)},\tilde{\rho}_{i}^{(1,2)},\tilde{\rho}_{i}%
^{2,1},\tilde{\rho}_{i}^{(3,1)},\tilde{\rho}_{i}^{(3,2)}%
\]
with $\dim\tilde{\rho}_{i}^{(d,l)}=d$ and the index $l$ counting different irreps
of the same dimension; there exists then a transformation $u$
such that
\[
u^{\dagger}\rho_{i}u=\left(  \bigoplus_{l=1}^{2}\tilde{\rho}_{i}
^{(1,l)}\right)  \oplus\left(  I_{2}\otimes\tilde{\rho}_{i}^{(2,1)}\right)
\oplus\left(  \bigoplus_{l=1}^{2}\left(  I_{3}\otimes\tilde{\rho}%
_{i}^{(3,l)}\right)  \right)  \,.
\]
The connectivity matrices $\rho_{i}$ are thus of the form as discussed in
Section \ref{sec:sec-part}. Note that the sub-matrices $S_{d,l}^{G}$
related to $d$-dimensional irreps occur now $d$ times in the decomposition. We
thus have 5 independent sub-spectra, 2 of dimension 24, 1 of dimension 48 and
2 of dimension 72 of which the latter are of multiplicity two and three,
respectively. The huge degeneracy in the spectra can clearly be seen in the
spectral statistics; it is manifest in the peak a $s=0$ in $P(s)$ 
(see Figure \ref{fig:stat}b) and leads to
\[
\begin{tabular}
[c]{ccc}
$K(\tau)=(2\cdot3^{3}+1\cdot2^{3}+2\cdot1^{3})=8/3$ & for & $\tau>3/24\,.$%
\end{tabular}
\ \
\]
The spectra appear to be uncorrelated otherwise; note however, that the
spectrum for each sub-matrix $S_{d,l}^{G}$ alone are correlated following $CUE$
statistics, which manifests itself in the deviations from purely Poisson
behaviour in $P(s)$ (cf. dashed curve) as well as in the behaviour of the form
factor for $\tau\leq 3/24$ which is dominated by the sub-spectra of the three
dimensional irreps.

\subsection{The generic case: no symmetries}

The overwhelming number of decompositions of the form (\ref{cond-Jd}) will of
course have no common symmetry apart from the trivial symmetry discussed in
section \ref{sec:sec-part}. Even though no further analytical results can be
given in this case, a numerical study may reveal interesting insights into the
range of validity of the universal RMT - regime. Figure \ref{fig:stat}c)
shows the level statistics of a regular quantum graph obtained from a fully
connected graph for a generic choice of connectivity matrices. One indeed
finds good agreement with random matrix theory for the CUE - ensemble. Deviations 
in the formfactor for small $\tau$ can be
attributed to the fact that the spectrum of $C$ is contained in the full
spectrum. After removing this separable part of the spectrum as done for the
NNS in Figure \ref{fig:stat}c) there is good agreement with random matrix results.
It is worth keeping in mind, that this is a highly non-random
matrix; we are dealing here with the spectrum of the $(n^{2}\times n^{2})$ unitary 
matrix $S^G$ which has only $n^{3}$ non-zero elements of which only $n^{2}$ are 
independent. In particular, the arc lengths in the graph are not incommensurate, 
the $n^{2}$ arcs in the graph share indeed only $n$ different lengths among them. 
Still, universality is obtained. The origin of the complexity in this type of quantum
graphs is here clearly not due to the 'randomness' in the choice of the matrix
elements but due to \emph{the lack of a common symmetry} in the set of connectivity 
matrices.

\section{Regular de Bruijn quantum graphs \label{sec:rmt}}
The results in the last section suggest that spectral statistics
of regular quantum graphs can to a large extent be controlled 
by properties of the permutation matrices $\rho_i$ independently 
of the coin $C$. It is thus natural to ask whether we can 
reduce the dimension of the coin to its smallest possible value, 
namely $\dim C = 2$, by considering large 2 - regular quantum graphs 
and still obtain random matrix correlations.

We can only expect random matrix statistics on a regular quantum
graph if the corresponding quantum graph with randomly chosen 
arc-lengths falls into the random matrix category. We therefore
need to consider 2-regular graphs leading to fast
(classical) mixing and not for example diffusive networks 
like ring graphs \cite{SS00} exhibiting 1d-Anderson localisation. 
The d-regular digraphs with the fastest mixing rates are so-called
$d$-ary \emph{de Bruijn graphs} of order $k$, $B(d,k)$, being 
the $k-1$st line graph generation of a complete graph of size $d$. That
is 
\[ B(d,k) = L^{k-1} G \quad \mbox{with} \quad A^G = J_d \]
and $L^k G$ is iteratively defined as $L^k G = L(L^{k-1} G)$ 
\cite{B46,FYA84}. De Bruijn graphs have size $d^k$ and are 
equipped with a complete symbolic dynamics of order $d$.
They play an important role in coding theory and parallel
algorithms \cite{SR91,SP89}. Numerical evidence suggests that 
quantum graphs based on de Bruijn digraphs with incommensurate arc 
lengths follow random matrix statistics in the limit of large graphs 
sizes even for $d=2$ \cite{Ta00,Ta01}.

\begin{figure}
  \begin{center} \includegraphics[height=5cm]{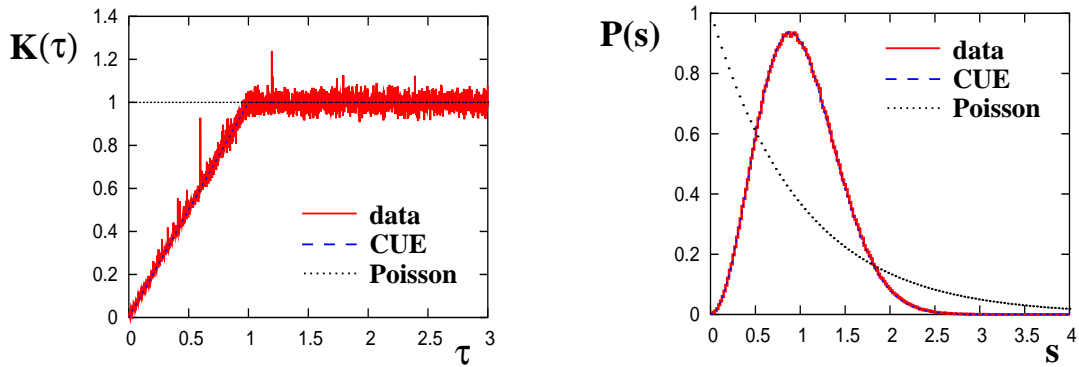} \end{center}
  \caption{Formfactor $K(\tau)$ and nearest neighbour spacing
  distribution $P(s)$ for binary de Bruijn graph of order $k=9$ with 
  $\dim S^{B(2,9)} = 1024$.} 
  \label{fig:deBrstat} 
\end{figure}

In Fig.\ \ref{fig:deBrstat} we show results for a regular quantum 
graph based on a binary de Bruijn graph $B(2,k)$ with k = 9 and 
quantum propagator
\[ 
S^{B(2,k)} = \left(\rho_1 \oplus \rho_2 \right) \cdot 
\left(C\otimes I_{2^k}\right)
\]
where $\rho_1, \rho_2$ are permutation matrices with 
$\rho_1+\rho_2 = A^{B(2,k)}$ and $C$ is a $(2 \times 2)$ unitary
matrix. The statistics in Fig.\ \ref{fig:deBrstat} are obtained
by averaging over the space of $(2\times 2)$ unitaries with 
respect to the Haar measure. The connectivity matrices of
dimension $2^k$ have been chosen randomly. To avoid accidental
symmetries, different sets of connectivity matrices have been 
produced and the statistics of the corresponding ensemble
averages combined. The spectral statistics of these regular 
quantum graphs agrees again very well with CUE statistics.
Recall that the unitary matrix $S^{B(2,9)}$ of size 
$2^{10}$ = 1024 has $2^{11}$ nonzero matrix elements 
which do, however, take on only 4 different (complex) values!
We have thus constructed extremely non-random matrices which still
show universal random matrix statistics and have thereby shown that
the BGS - conjecture is valid far beyond regimes previously thought
to be included in the conjecture. 
Similar numerical results were also found for the spectra of the Laplacian 
of regular graphs \cite{JMRR99}. A Laplacian on a $d$ - regular undirected, 
loop-less graph $G$ is $\Delta = d I_n - A^G$ and is thus a symmetric 
matrix whose non-zero matrix elements take on the values 1 or $d$ only. When
averaging over sets of $d$ - regular graphs agreement with GOE statistics 
was found. This underlines once more that the origin of universality in 
spectral statistics lies not in the randomness of the matrix elements.

\section{Conclusions}
We introduce a decomposition of certain regular quantum graphs
which separates the quantum propagator on a graph into
a topological part containing the connectivity matrices and a 
trivial part containing the quantum scattering information at 
the vertices. This allows one to implement global symmetries on the graph
by choosing the connectivity matrices according to desired symmetry 
properties. We demonstrate that the complexity in the quantum spectrum
(which may be seen to take on its maximal value when the statistics 
coincides with RMT) can here be linked to the amount of complexity 
contained in the set of permutation matrices building up the quantum graph. 
We present examples, where for a given graph and a fixed coin
matrix, we were able to construct anything from Poisson to RMT-statistics 
just by changing the set of connectivity matrices. By doing so, we leave 
the local properties of the graph invariant, but change the way in 
which incoming and outgoing channels between vertices are connected 
and thus the global structure of the wave dynamics. We take this concept
to its extreme by demonstrating numerically that unitary matrices 
representing 2-regular quantum graphs whose non-zero matrix elements 
take on only four different values still follow CUE statistics
for de Bruijn graphs.

We believe that our results open up new perspectives in understanding 
universality in spectral statistics. It transforms the question 
from a continuous into an essentially discrete problem focusing on 
the way local scattering processes are connected and condensing the 
parameter space to an absolute minimum, (namely four-dimensional).\\

\noindent
\textbf{Acknowledgement:} 

\noindent
We would like to thank Andreas Winter for interesting discussions and 
Stephen Creagh and Jens Marklof for reading the manuscript and 
for valuable comments. The main part of the work was carried 
out during a {\em Royal Society Industrial Fellowship} by one of us 
(GT) who would like to thank the Royal Society and Hewlett-Packard, 
Bristol, for financial support.


\begin{thebibliography}{999999}                  

\bibitem[B46]{B46} N.~G.~de Bruijn, 
\emph{Konink.  Nederl. Akad. Wetersh. Verh. Afd. Naturk. Eerste Reelss}, 
\textbf{A49}, 758-764 (1946).

\bibitem[BG01]{BG01}J.~Bang-Jensen and G.~Gutin, \emph{Digraphs. Theory,
algorithms and applications}, Springer Monographs in Mathematics,
Springer-Verlag, London, 2001.
 
\bibitem[BGS84]{BGS84} O.~Bohigas, M.J.~Giannoni and C.~Schmit, 
\emph{Phys.~Rev.~Lett.} \textbf{52}:1 (1984).
 
\bibitem[BK99]{BK99} G.~Berkolaiko and J.~P.~Keating 
{\em J.\ Phys.\ A} {\bf 32}, 7827-7814 (1999).

\bibitem[FYA84]{FYA84} M.~A.~Fiol, J.~L.~A.~Yebra and I.~Alegre, 
\emph{IEEE Trans.~Comput.} \textbf{33}, 400-403 (1984).

\bibitem[JMRR99]{JMRR99} D.~Jakobson,  S.~Miller,  I.~Rivin and  Z.~Rudnick'  
IMA Mathematics and its Applications \textbf{109}, 317-329 (1999).
 
\bibitem[Ke03]{Ke03}J.~Kempe, \emph{Contemporary Physics}, \textbf{44} (4),
307-327 (2003). quant-ph/0303081.
 
\bibitem[KS97]{KS97}T.~Kottos and U~Smilansky, \emph{Phys.~Rev.~Lett.}
\textbf{79} 4794 (1997).
 
\bibitem[KS99]{KS99}T.~Kottos and U.~Smilansky, \emph{Ann.~Phys.~NY
}\textbf{274} 76 (1999). chao-dyn/9904007.
 
\bibitem[Ku02]{Ku02}P.~Kuchment, \emph{Waves in Random Media}, \textbf{12}, R1 (2002).
 
\bibitem[PTZ03]{PTZ03}P.~Pako{\'{n}}ski P, G.~Tanner and K.~\.{Z}yczkowski,
\emph{J.\ Stat.\ Phys.} \textbf{111}, 1331-1351 (2003). nlin.CD/0110043.
 
\bibitem[S03]{S03}S.~Severini, \emph{SIAM J.~Matrix Anal.~Appl., }\textbf{25},
vol. 1, 295 (2003). math.CO/0205187.
 
\bibitem[S03a]{S03a}S.~Severini, math.CO/0309092.

\bibitem[SP89]{SP89} M.~R.~Samatham and D.~K.~Pradhan,
\emph{IEEE Trans.~Comput.} \textbf{38}, 567-581 (1989).

\bibitem[SR91]{SR91} M.~A.~Sridhar and C.~S.~Raghavendra
\emph{IEEE Trans.~Comput.} \textbf{40}, 1167-1174 (1991).

\bibitem[SS00]{SS00} H.~Schanz and U.~Smilansky 
{\em Phys.\ Rev.\ Lett.} {\bf 84}, 1427-1431 (2000).

\bibitem[ST04]{ST04}S.~Severini and G.~Tanner, in preparation, December (2003). 
 
\bibitem[Ta00]{Ta00}G.~Tanner, \emph{J. Phys. A: Math. Gen. }\textbf{33} 3567-3585
(2000). nlin.CD/0001025.
 
\bibitem[Ta01]{Ta01}G.~Tanner, \emph{J. Phys. A: Math. Gen.} \textbf{34} 8485-8500
(2001). nlin.CD/0104014.
\bibitem[Te03]{Te03}A.~Terras, \emph{Amer.\ Math.\ Monthly} \textbf{109},
 121 - 139 (2003); A.~Terras, {\em Fourier Analysis on Finite Groups and Applications}, 
(Cambridge University Press, Cambridge, 1999).
 
\end{thebibliography}
\end{document}